# First-principles modeling of oxygen interaction with SrTiO$_3$(001) surface: Comparative density-functional LCAO and plane-wave study

Vitaly Alexandrov[1], Sergei Piskunov[2,3,4]*, Yuri F. Zhukovskii[4], Eugene A. Kotomin[4,5] and Joachim Maier[5]

[1]Department of Chemical Engineering and Materials Science and NEAT ORU, University of California, 1 Shields Avenue, Davis, California 95616-5294, USA
[2]Faculty of Computing, University of Latvia, 19 Raina blvd., Riga LV-1586, LATVIA
[3]Faculty of Physics and Mathematics, University of Latvia, 8 Zellu Str., Riga LV-1002, LATVIA
[4]Institute of Solid State Physics, University of Latvia, 8 Kengaraga Str, Riga LV-1063, LATVIA
[5]Max-Planck-Institute for Solid State Research, Heisenbergsrtr. 1, Stuttgart 70569, GERMANY

**ABSTRACT**

Large scale first-principles calculations based on density functional theory (DFT) employing two different methods (atomic orbitals and plane wave basis sets) were used to study the energetics, geometry, the electronic charge redistribution and migration for adsorbed atomic and molecular oxygen on defect-free SrTiO$_3$(001) surfaces (both SrO- and TiO$_2$-terminated), which serves as a prototype for many ABO$_3$-type perovskites. Both methods predict substantial binding energies for atomic O adsorption at the bridge position between the oxygen surface ions and an adjacent metal ion. A strong chemisorption is caused by formation of a surface molecular peroxide ion. In contrast, the *neutral molecular* adsorption energy is much smaller, ~0.25 eV. Dissociative molecular adsorption is energetically not favourable, even at 0 K. Adsorbed O atoms migrate along the (001) direction with an activation energy of ~1 eV which is much larger than that for surface oxygen vacancies (0.14 eV). Therefore, the surface O vacancies control encounter with the adsorbed O atoms and oxygen penetration to the surface which is the limiting step for many applications of ABO$_3$-type perovskites, including resistive oxygen sensors, permeation ceramic membranes and fuel cell technology.

**INTRODUCTION**

The interaction of gas-phase oxygen with the perovskite surfaces is important for the fundamental understanding of oxide reactivity and for many applications involving permeation membrane technology, sensors, capacitors, photocatalysts, as well as fuel cell electrodes [1,2]. The difficulties of distinguishing between adsorbed and lattice oxygen (unless the adsorbed species carry a net spin) result in a significant lack of experimental information about the oxygen adsorption processes occurring on perovskite surfaces. Also, the *ab initio* simulations of oxygen adsorption on ABO$_3$ perovskites are very scarce in the literature. Recently, we reported first principles calculations for surfaces of Sr-doped LaMnO$_3$ [3] used as fuel cell cathodes.

In this paper, we present results of large-scale calculations of adsorption and diffusion of atomic and molecular oxygen on the (001) surface of a cubic SrTiO$_3$ which serves as a prototype for a wide class of ABO$_3$ perovskites. We use two different versions of the first principles DFT approaches which are discussed below. The main questions that we want to answer in our comparative study were the following: (*i*) which adsorption sites are most energetically favorable, (*ii*) what is the nature of bonding of adsorbed oxygen atoms and molecules to the

SrTiO₃ surfaces, (*iii*) what is the diffusion mechanism of the adsorbed species, (iv) what is the difference in mobilities of adsorbed oxygens and surface vacancies, (v) how a choice of the two different calculation methods affects the results.

**Computational details**

In our comparative study we have used two quite different DFT approaches: the *hybrid* exchange-correlation functional with optimized LCAO (linear combination of atomic orbitals) basis set (BS) and the generalized gradient approximation (GGA) with plane-wave (PW) BS. To perform hybrid calculations, we have used the *CRYSTAL* computer code (see [4] and references therein). This code employs the Gaussian-type functions for an expansion of the crystalline orbitals. The BSs used in this study were carefully optimized by us elsewhere [5]: O—8-411(1d)G, Ti—411(311d)G, and Sr—311(1d)G, whereby inner core electrons of Sr and Ti atoms are described by Hay-Wadt effective core potentials [6]. The exchange-correlation B3PW functional [7] involves a hybrid of non-local HF exact exchange and exchange potentials constructed using both Local Density Approximation (LDA) and non-local GGA, combined with the GGA correlation potential of Perdew and Wang. Our previous studies show that the B3PW exchange-correlation technique is a reliable tool to calculate both surface properties and electronic structure of defective ABO₃ perovskites [8]. This gives us a firm ground to adopt the B3PW hybrid functional in the present study of oxygen adsorption on the SrTiO₃ (001) surfaces. Preliminary *CRYSTAL* calculations were discussed in Ref. [9].

In our PW simulations on SrTiO₃(001) surfaces, we have used the *VASP 4.6* code [10], with the PBE-GGA nonlocal functional [11] employed for both exchange and correlation. The cut-off energy was chosen to be 400 eV. The vacuum gap along the z axis was chosen 19.5 Å.

We considered both SrO- and TiO₂-terminated SrTiO₃(001) surfaces. These were modeled using 7-layer symmetrically terminated slab unit cells [12]. In order to simulate the isolated adsorbed species, the aforementioned slab unit cell was extended in *CRYSTAL* calculations to a 2×2 slab surface supercell. The reciprocal space integration was performed by sampling of the Brillouin zone (BZ) of the supercell with the 6×6×1 Pack-Monkhorst net [13] which provides a balanced summation over the direct and reciprocal lattices. In the VASP calculations $2\sqrt{2} \times 2\sqrt{2}$ primitive unit cells corresponding to 12.5% surface defect coverage and a 2×2×1 Monkhorst-Pack mesh of *k* points [13] was used. The initial guess for relaxation of bare SrTiO₃(001) surfaces was taken from [12], further optimization was performed using analytical methods as implemented in both *VASP* and *CRYSTAL* codes (including nudged elastic band method (NEB) in the VASP [4]). To save the computational time, in the LCAO calculations of O adsorbed atop surface Ti and Sr ions, the O-Me distances were taken the same as obtained in the PW calculations (but all other ions were allowed to relax).

In order to obtain an overall picture of the equilibrium geometry, energetics, and the electronic structure, we have studied the adsorption of atomic oxygen at a variety of sites on the relaxed but unreconstructed cubic SrO- and TiO₂-terminated SrTiO₃(001) surfaces as shown at Figure 1. Oxygen atom adsorption energies were calculated with respect to the energy of an oxygen free atom in the triplet ground state:

$$E_{ads}^{(at)} = -\frac{1}{2}\left(E_{tot}^{system} - E_{tot}^{slab} - 2E_{tot}^{O_{triplet}}\right). \tag{1}$$

where $E_{tot}^{slab}$ is the total energy of fully relaxed substrate slab, $E_{tot}^{O_{triplet}}$ is the energy of an isolated oxygen atom in the ground triplet state, and $E_{tot}^{system}$ is the energy of a slab containing an adsorbate. The prefactors ½ and co-factor of 2 before $E_{tot}^{O_{triplet}}$ appear since the interface is modeled by a substrate slab with two equivalent surfaces and two $O_{ads}$ atoms symmetrically positioned on both sides of slab. If $E_{ads}$ is positive, the adsorption is energetically favorable.

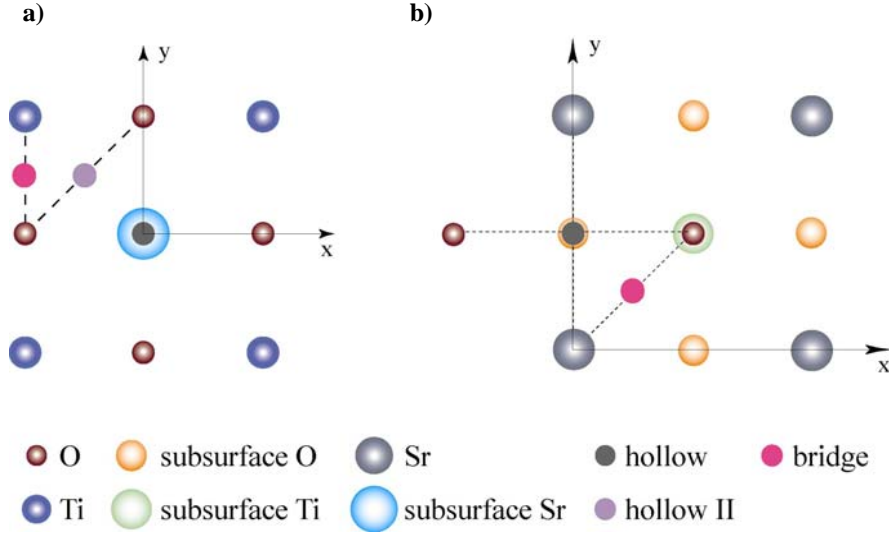

**Figure 1**. Top views of the possible positions for oxygen adsorption on (a) $TiO_2$- and (b) SrO-terminated (001) substrate (see labels).

## RESULTS AND DISCUSSION

### Adsorption of atomic oxygen

Table I comprises the calculated O atom adsorption energies (with respect to a free O atom in the triplet state) for all five surface sites under consideration (Fig. 1). Both methods predict the most favorable configurations of oxygen adsorption on both surface terminations to be the *bridge position*. The calculated binding energies are typically larger in the PW method.

The analysis of the interatomic distances (Table II) indicates that at the bridge position the adatom is substantially shifted towards the nearest surface metal atom. A further analysis by means of the effective atomic charges (Bader and Mulliken approaches, Table III) along with the electron density maps (Fig. 2) points at the formation of pseudo-molecular tilted species $O_{ads}$-$O_{surf}$ (peroxide $O_2^{2-}$ ion).

The redistribution of the electron density in the vicinity of the adsorbed oxygen atom (Fig. 2) was obtained within the hybrid HF-DFT method, while it is known that the pure DFT approach commonly overestimates the covalent contribution in the chemical bonding. This also explains the generally larger adsorption energies obtained within the pure DFT method (PW-PBE) (Table I).

Nevertheless, both methods draw the same conclusions about the atomic oxygen adsorption on the surfaces of $SrTiO_3$ crystal. Specifically, (*i*) there is a strong chemisorption with the highest adsorption energy for the bridge position; (*ii*) $O_{ads}$ is bound stronger on the more ionic SrO surface than on the $TiO_2$ termination.

**Table I.** The oxygen adsorption energies (in eV) from eq.(1) calculated using the PW-PBE method (VASP code) and the LCAO-B3PW (CRYSTAL code). The adsorbed oxygen atom at the hollow sites was considered in the ground (triplet) state. The energetically most favorable positions are given in bold.

| Site | PW-PBE | LCAO-B3PW | Site | PW-PBE | LCAO-B3PW |
|---|---|---|---|---|---|
| $TiO_2$-termination | | | SrO-termination | | |
| Ti | 2.13 | 0.70 | Sr | 0.57 | 0.37 |
| O | 2.51 | 1.76 | O | 2.44 | 1.54 |
| Bridge | **2.96** | **2.03** | Bridge | **3.06** | **2.43** |
| Hollow | 1.12 | 0.93 | Hollow | 1.73 | 1.08 |

**Table II.** Distances (in Å) between the adsorbed oxygen atom and the nearest $O_s$ and $Ti_s$ (or $Sr_s$) atoms for optimized adsorption structures of $SrTiO_3$ (001) substrates.

| Site | PW-PBE | | LCAO-B3PW | |
|---|---|---|---|---|
| | $O_{surf}$ | $Ti(Sr)_{surf}$ | $O_{surf}$ | $Ti(Sr)_{surf}$ |
| $TiO_2$-termination | | | | |
| Ti | 2.45 | 1.65 | 2.45 | 1.65 |
| O | 1.45 | 2.31 | 1.46 | 2.76 |
| Bridge | 1.46 | 1.56 | 1.47 | 1.91 |
| Hollow | 2.43 | 3.18 | 2.24 | 2.00 |
| SrO-termination | | | | |
| Sr | 3.32 | 2.33 | 3.32 | 2.33 |
| O | 1.47 | 2.85 | 1.47 | 3.24 |
| Bridge | 1.49 | 2.09 | 1.50 | 2.45 |
| Hollow | 2.62 | 2.47 | 2.48 | 2.53 |

**Table III.** The effective atomic charges, in $e$ (Bader analysis in PW-PBE and Mulliken analysis in LCAO-B3PW), for the adsorbed oxygen atom and the nearest surface O and Ti (or Sr) atoms for optimized adsorption structures of $SrTiO_3$ (001) substrates. Atomic charges at the defect-free surface layer are: PW-PBE: Ti 2.03, O -1.16e ($TiO_2$ termination); Sr 1.56, O -1.28e (SrO termination); LCAO-B3PW: Ti 2.31, O -1.32e; Sr 1.84, O -1.52e.

| Site | PW-PBE | | | LCAO-B3PW | | |
|---|---|---|---|---|---|---|
| | $O_{ads}$ | $O_{surf}$ | $Ti(Sr)_{surf}$ | $O_{ads}$ | $O_{surf}$ | $Ti(Sr)_{surf}$ |
| $TiO_2$-termination | | | | | | |
| Ti | -0.76 | -1.12 | 2.03 | -0.60 | -1.10 | 2.39 |
| O | -0.61 | -0.69 | 2.05 | -0.62 | -0.77 | 2.29 |
| Bridge | -0.49 | -0.71 | 2.13 | -0.52 | -0.79 | 2.28 |
| Hollow | -0.23 | -1.13 | 2.02 | -0.29 | -1.16 | 2.29 |
| SrO-termination | | | | | | |
| Sr | -0.71 | -1.22 | 1.54 | -0.65 | -1.32 | 1.87 |
| O | -0.73 | -0.80 | 1.58 | -0.71 | -0.90 | 1.86 |
| Bridge | -0.79 | -1.07 | 1.56 | -0.84 | -0.88 | 1.85 |
| Hollow | -0.72 | -1.12 | 1.59 | -0.52 | -1.32 | 1.86 |

In both methods the energy gain due to adsorption of two oxygen atoms is smaller than the free molecule dissociation energy (experimental value 5.12 eV [16], cf. with calculated values of 5.30 eV LCAO and 6.05 eV PW-PBE). Therefore, dissociation of oxygen molecule due to adsorption on a defect-free surface is energetically unfavorable.

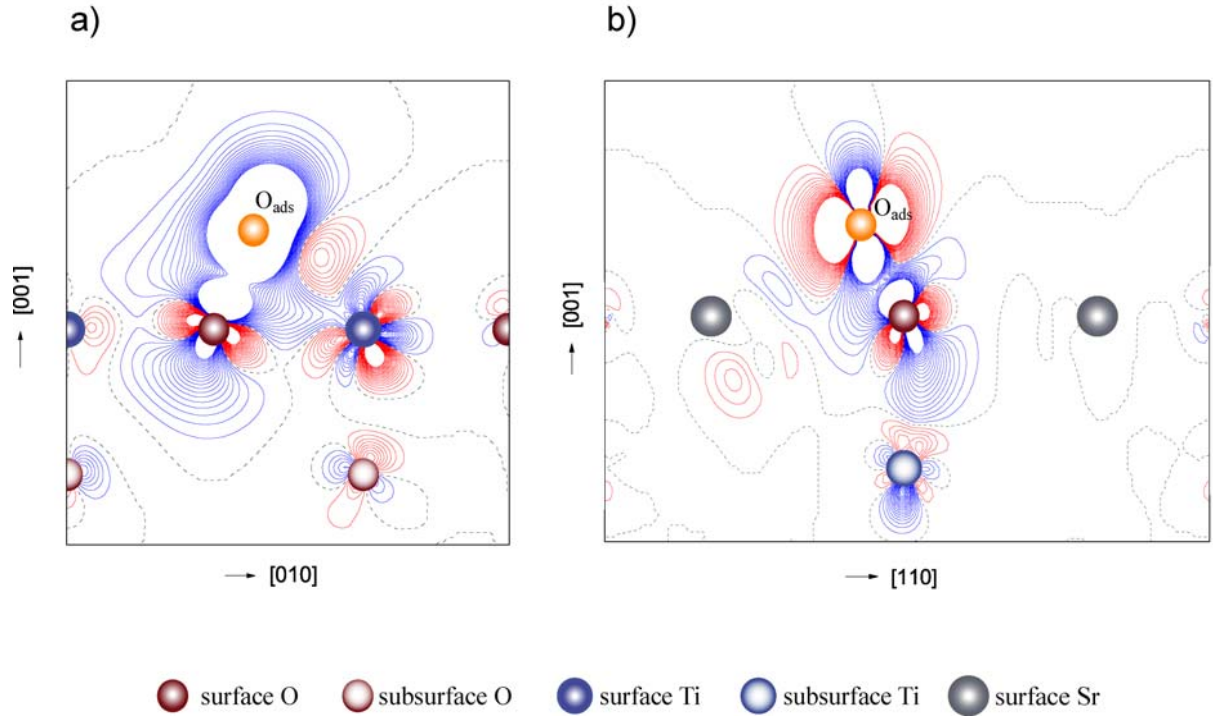

**Figure 2**. The differential electron density maps (the total density of the surface containing an adsorbate minus the sum of the electron densities of both isolated oxygen atom and bare slab) for $O_{ads}$ over the bridge position on (a) $TiO_2$-terminated and (b) SrO-terminated surfaces. On the electron density map, solid (red), dash (blue) and dash-dot (black) lines describe positive, negative and zero values of the induced electron density, respectively.

**Adsorption of molecular oxygen**

We explored a number of possible sites and orientations for the adsorption of an $O_2$ molecule on both terminations of $SrTiO_3$(001). (Note that the LCAO-calculated bond lengths of a free $O_2$ molecule are 1.20 Å and 1.23 Å for LCAO and PW methods, in good agreement with experiment, 1.21 Å [16]). The LCAO calculations show that molecular adsorption occurs at hollow sites (hollow II for $TiO_2$ termination as shown in Fig. 1) with the molecular axis oriented perpendicular to the surface. The relevant adsorption energies $E_{ads}$ are 0.02 and 0.09 eV for SrO- and $TiO_2$- terminated $SrTiO_3$(001) surfaces, respectively.

In turn, the PW-PBE method yields the largest $E_{ads}$=0.25 eV for the bridge position between Ti and O surface atoms at the $TiO_2$-terminated surface (the molecular axis is just slightly tilted from normal vector toward surface O atom). In both the LCAO and the PW-PBE methods the equilibrium distance between surface and molecular adsorbate is found to be as large as ~3 Å.

The Mulliken population analysis shows practically no charge transfer between the surface atoms and adsorbate. This fact and invariance of the $O_2$ bond length, demonstrate that we are dealing with quite weak physisorption.

**Diffusion of oxygen species**

The limiting step for many perovskite applications (e.g. fuel cells and permeation membranes) is the penetration of the adsorbed oxygen atom into the nearby surface oxygen vacancy. The kinetics of this process is controlled by mutual encounter of the two species and thus needs calculation of the migration energies of both adsorbed O atoms and oxygen vacancies. First of all, it follows from the Table 1 that the migration energy of the adsorbed oxygen between the most energetically favorable bridge positions along the (100) direction via the position above Ti ions on the $TiO_2$-terminated surface requires 0.83 eV according to the *VASP* and 1.33 eV in the *CRYSTAL* calculations. (The main discrepancy stems from the low binding energy predicted for O atop Ti in the latter calculations, the reason of which is not completely clear.)

On the SrO terminated surface, the migration occurs along the (110) directions with the hollow position as the saddle point. The relevant energies are very close, 1.33 eV and 1.35 eV for *VASP* and *CRYSTAL*, respectively. These energies are larger than those for $TiO_2$ termination due to a larger O binding energy resulting from more negative O ions on the SrO surface (Table 3). Recent VASP calculations [17] show the same trend, 0.81 eV and 0.68 eV for SrO and $TiO_2$ terminations, though the absolute values which differ from our study, probably due to larger supercells used there.

Another step was the calculation of the O *vacancy* migration energy along the $TiO_2$ (001) surface using the VASP code. The value of the barrier is expected to be lower than for the bulk because of the reduced coordination number for surface atoms. Our calculations show that the barrier is indeed 0.14 eV, being almost by a factor three lower than that calculated for the bulk (0.38 eV). Note that the bulk value is too low compared to the experimental data (0.86 eV [18]). The reason could be that we treated a neutral vacancy (electron density from missing O *atom* is trapped by two nearest Ti ions) whereas experiments refer to the charged vacancies at high temperatures (electrons trapped somewhere in crystal far from a vacancy). Our energies are similar to another VASP study [19] (see also review article [8]). Thus, we predict highly mobile surface oxygen vacancies as compared to the vacancies in the bulk as well as adsorbed O atoms.

Lastly, computer simulations of the adsorbed oxygen atom dropping into the closest oxygen vacancy reveal a distinguishable but extremely small activation barrier, just 0.01 eV. Hence, we predict an almost no-barrier penetration of the adsorbed O atom into the nearby vacancy that occurs when the strongly-bound adatom meets the very mobile surface oxygen vacancy.

**CONCLUSIONS**

We have performed a number of DFT-based calculations on a cubic $SrTiO_3(001)$ substrate containing oxygen adsorbates. The most favorable adsorption sites for oxygen atoms are bridge positions on both terminations. On the SrO-terminated substrates, the oxygen adatom is bound more strongly than on $TiO_2$-terminated one. Our calculations show that the $O_{ads}$ atoms tend to form the molecular peroxide ions with surface oxygens which is consistent with a recent VASP study [17]. In contrast, on a clean defect-free $SrTiO_3(001)$ substrate, molecular adsorption is very weak.

The diffusion of atomic oxygen along the surfaces is expected to have relatively high activation energies and to be quite anisotropic. We predict almost barrier-free penetration of the adsorbed O ions into the much more mobile surface vacancies, similarly to what we found recently for $LaMnO_3$ [3,15]. This demonstrates that a high concentrations of mobile vacancies are important for high performance of ceramic membranes for gas separation or for fuel cell electrodes. A DFT study of more complex perovskite materials such as $(Ba,Sr)(Co,Fe)O_3$ (BSCF) is in progress.

## ACKNOWLEDGEMENTS

Many thanks to R. Merkle, E. Heifets and Yu. Mastrikov for numerous stimulating discussions. This study has been partly supported through the European Social Fund (ESF) project Nr. 2009/0216/1DP/1.1.1.2.0/09/APIA/VIAA/044 and EC FP7 NASA-OTM project. VA thanks the Max Planck Institute for Solid State Research and its International School for Advanced Materials for financial support and hospitality during his PhD study.